\documentstyle[aps,prl,multicol,epsfig]{revtex}
\begin{document}
\draft
\title{
Landauer-like formula for dissipative tunneling}
\author{
S.G. Chung
}
\address{
Max-Planck-Institut f\"{u}r Physik komplexer Systeme, N\"{o}thnizer
Str. 38, D-01187 Dresden, Germany
}

\maketitle

\begin{abstract}
The Landauer formula for electrical conductance is simple but works remarkably
well in mesoscopic systems.  We propose a Landauer-like formula for
calculating an escape rate out of a dissipative metastable well, the quantum
Kramers rate.
The proposed formula works well in the current-voltage characteristic of a
single Josephson junction.
\end{abstract}



\begin{multicols}{2}

The electrical conduction is a typical non-equilibrium phenomenon.  Two
theories are known.  One is the Green-Kubo-Nakano linear response theory\cite{Kubo} and
the other is the Landauer theory\cite{Landauer}.  The former has a generality
but often technically challenging.  The latter, on the other hand, lacks a
generality and limited to mesoscopic systems so far such as quantum wires, but
is intuitive, simple and practically quite useful.  

The system Landauer considered  is composed of a sample which is connected by
two leads to the electron reservors with chemical potentials $\mu_{1,2}$.
The voltage applied is $V=-(\mu_{1}-\mu_{2})/e$.  The electrons which
contribute to the net current are those electrons with energies between
$\mu_{1}$ and $\mu_{2}$.  Write the energy of electrons with wave number $k$
as $E(k)$, then it contributes $-e(dE/\hbar dk) T(k)$ where
$T(k)$ denotes the transmission rate.  Assuming that any electron supply and
removal from resevors are immediate, the net current is simply, including
factor 2 for up and down spins,

\begin{equation} \label{eq1}
I=-2e\int_{k_1}^{k_2}\frac{dk}{2\pi}
\frac{dE}{\hbar dk}T(k)
\end{equation}
where $\mu_{1,2}=E(k_{1,2})$.  When $\mu_1\sim\mu_2$ and $T(k)=const=T$,
one gets the conductance
\begin{equation} \label{eq2}
G=\frac{I}{V}=\frac{2e^2}{h}T
\end{equation}

The purpose of the present paper is to propose a Landauer-like formula for
calculating a particle's escape rate out of a dissipative metastable well, the
so-called quantum Kramers rate.  This well-defined and universal problem is
ubiquitous in quite a wide range of physics and attracted much attention over
decades\cite{Hanggi,Mel,Weiss,Thorwart,Dittrich}.
The latest review article discussed a seemingly ultimate issue of full
dynamical analysis in the presence of time-dependent external driving
force\cite{Thorwart}.  In spite of all these quite intensive studies, however, the
issue appears to be still open.  The situation looks somehow similar to the
problem of electrical conductance.  As mentioned above, we have the linear
response theory which is exact but often technically challenging and the
Landauer theory which is limited but intuitive, simple and has seen a remarkable
success in mesoscopic systems.  It may be worth pursuing an analogous simple
formula for the quantum Kramers rate.  

First note that (\ref{eq1}) is a sum of individual contributions
$(dE/\hbar dk)T(k)$ which is nothing but a right going flux.  A
corresponding quantity in the escape rate is the probability of
finding particle moving rightward at the top of the potential barrier.  Adding
all such contributions, we might reach a formula
\begin{equation} \label{eq3}
v_{qts}=Tr\left[e^{-\beta H}{\cal F} \Theta(p)\right]/Z_0
\equiv \frac{1}{2M}<\delta(x)|p|>
\end{equation}
where $Z_0$ is the partition function of the metastable well, 
${\cal F}\equiv \delta(x)p/M$ with $p=-i\hbar \partial_x$ is the
flux at the barrier top and $\Theta (p)$ is the unit step function.
The formula (\ref{eq3}) is nothing but the quantum transient state (QTS) theory
long known in chemistry for calculating molecular reaction rates\cite{Miller}.
Historically, the QTS theory first appeared as an approximation to a
formally exact expression from scattering theory\cite{Yamamoto},
\begin{equation} \label{eq4}
v_{exact}=\rm Re <{\cal FP}>
\end{equation}
where Re means to take the real part and the operator
\begin{equation} \label{eq5}
{\cal P} 
=
\lim_{t \to \infty} 
e^{iHt/\hbar}\Theta (p) e^{-iHt/\hbar}
\end{equation}
projects onto all states that have positive momentum in the infinite future.
The exact expression (\ref{eq4}) can be manipurated to give\cite{Yamamoto}
\begin{equation} \label{eq6}
v_{exact}
=\frac{1}{2}\int_{-\infty}^{\infty} dt <{\cal F} (0) {\cal F} (t)>
\end{equation}
where ${\cal F} (t)$ is the Heisenberg picture of the operator ${\cal F}$.  

The formula (\ref{eq6}) takes precisely the form of the Green-Kubo-Nakano
formula for the electrical conductance.  The QTS theory was obtained from the
exact one (\ref{eq4}) by simply replacing ${\cal P} \to \Theta (p)$ thereby
neglecting all the dynamical processes.  The QTS formula (\ref{eq3}), however,
is not well-defined quantum mechanically because the operator ${\cal F}$ is ambiguous.
The necessary regularization of such operators as ${\cal F}$ is done by the
Weyle rule\cite{Weiss}.  Let us write (\ref{eq3}) as 
\begin{eqnarray} \label{eq7}
v_{qts}
&=&
\frac{1}{2MZ_0}\int \int dxdx'
\nonumber\\
&\times&
<x|e^{-\beta H}|x'>
<x'|\delta(x)|p||x>
\end{eqnarray}
Now the Weyle rule gives a quantum operator from a classical one, 
${\cal F}_{cl} (p,q) \to {\cal F}_{op} (p,q)$, in coordinate matrix
representation\cite{Miller},
\begin{equation} \label{eq8}
<x'|{\cal F}_{op}|x>
=
\frac{1}{h}
\int dp e^{-ip(x-x')/\hbar} 
{\cal F}_{cl} 
\left(
p,\frac{x+x'}{2}
\right)
\end{equation}
Applied to the last term in (\ref{eq7}), we have
\begin{eqnarray} \label{eq9}
<x'|\delta(x)|p||x>
&=&
\frac{1}{h}
\int dp  
e^{-ip(x-x')/\hbar}
|p|
\delta
\left(
\frac{x+x'}{2}
\right)
\nonumber\\
&=&
\frac{\hbar}{\pi}
\delta
\left(
\frac{x+x'}{2}
\right)
\partial_x 
\left(
\frac{1}{x-x'}
\right)
\end{eqnarray}
Putting (\ref{eq9}) into (\ref{eq7}) and integrating in parts gives
\begin{equation} \label{eq10}
v_{qts}
=
\frac{\hbar}{4\pi MZ_0}\int dx (-\frac{1}{x})
\partial_x <x|e^{-\beta H} |-x>
\end{equation}

Some comments are due on the Weyle rule.  First, it is not exact.  For
example, if one proceeds with this rule for the propegator 
$e^{-iHt/\hbar}$ with $H=p^2/2M+V(x)$, the Weyle rule gives only a \textit{short
time approximation}\cite{Feynman}.  In general, as one can easily see from
(\ref{eq8}), the more complicated the functional forms of coordinate and
momentum, unless they are functions of coordinate alone or momentum alone, the
more dubious results are obtained by the Weyle rule.  Back to the propagator,
because it is already Hermite, the classical and quantum expression take the
same form.  Thus the failure of the Weyle rule resides in an inability of doing proper
commutation calculations for complicated mixtures of coordinate and momentum.
In fact when the starting classical form is simple enough, e.g. $xp$, then the
Weyle rule is known to give a correct answer, e.g. $(xp+px)/2$.  It is
certainly embarrasing that we still do not know how to construct correct
quantum mechanical operators from classical quantities when they are
complicated enough.  Second, a connection to
the more familiar Wigner distribution function should be mentioned.  In fact
we have for any operator ${\cal O}$,
\begin{equation} \label{eq11}
<{\cal O}>
=
\int\int dxdx'<x|e^{-\beta H}|x'><x'|{\cal O}_{op}|x>
\end{equation}
Using the Weyle rule (\ref{eq8}) for the last term in (\ref{eq11}) and
changing the variables as $x-x'=r,~\frac{x+x'}{2}=q$, we have
\begin{equation} \label{eq12}
<{\cal O}>
=
\int\int dpdq{\cal O}_{cl} (p,q) W(p,q)
\end{equation}
where the Wigner distribution function $W(p,q)$ is the Wigner representation
of the density matrix\cite{Weiss},
\begin{equation} \label{eq13}
W(p,q)
\equiv
\frac{1}{h}\int dr e^{-ipr/\hbar}
<q+\frac{r}{2}|e^{-\beta H}|q-\frac{r}{2}>
\end{equation}
The expression (\ref{eq12}) \textit{should not} be confused with the formally
exact expression
\begin{equation} \label{eq14}
<{\cal O}>
=
\int\int dpdq {\cal O}_W (p,q) W(p,q)
\end{equation}
where ${\cal O}_W$ is the Wigner representation of the operator ${\cal O}$,
\begin{equation} \label{eq15}
{\cal O}_W
=
\int dr e^{-ipr/\hbar}
<q+\frac{r}{2}|{\cal O}_{op}|q-\frac{r}{2}>
\end{equation}
It is a simple exercise to check that ${\cal O}_W$ reduces to ${\cal O}_{cl}$ if
the term in (\ref{eq15}) is \textit{approximated} by the Weyle rule (\ref{eq8}).
Third, as is seen in (\ref{eq9}) and (\ref{eq10}), the term
$\delta(q)=\delta(\frac{x+x'}{2})$ results in an antiperiodic paths
$<x|e^{-\beta H}|-x>$ in the expression (\ref{eq10}).  The nonlocality, an
integral over $x$ rather than just a contribution at the barrier top $x=0$, however, arises
from the step function $\Theta (p)$.  It is noted that some nonlocal
contribution is certainly negative reflecting the fact that the Wigner
distribution function is a quasi-probability function.  It is also
noted that the nonlocality does not show up when the QTS formula is treated by
a semiclassical approximation (see e.g. Eq(2.22) in the second paper
in\cite{Miller}).

We have critically reviewed the QTS theory (\ref{eq10}) as a candidate for the
Landauer-like formula for the quantum Kramers rate.  Two major approximations
involved are (A) static or transient-state approximation
and (B) the Weyle rule (\ref{eq8}).  The step function $\Theta (p)$ or
equivalently $|p|$ under the trace operation is found to be a \textit{dangerous}
operator for the Weyle procedure, leading to nonlocality and associated
possible negativity.  Our idea in this paper is to get rid of this dangerous
operator by taking \textit{square} and \textit{square root}.  We thus propose
a Landauer-like formula
\begin{equation} \label{eq16}
v_{Ll}
=
\frac{1}{2M}\sqrt{<\delta (x) p^2>/L}
\end{equation}
where $L$ is the size of the potential well.  The factor $L$ and the square
root operation arises dimensionally and the fact that ${\cal F}$ is a density
operator containing $\delta (x)$.  The proposed formula also makes the Kramers
rate calculable from the \textit{local}, at the barrier top, particle state.
The factor $1/2$ takes into account only half of the contribution, outgoing
one.  It may be worth emphasizing that the $p^2$ operator rather than $\Theta
(p)$ or $|p|$ would make the Weyle rule less dubious.  Now, proceeding as
before, we have
\begin{equation} \label{eq17}
v_{Ll}
=
\frac{\hbar}{2M}
\sqrt{
-\frac{1}{4LZ_0}\partial_{x}^2<x|e^{-\beta H}|-x>\arrowvert _{x=0}
}
\end{equation}
which should be compared with the QTS formula (\ref{eq10}).  

Note that the proposed formula (\ref{eq17}), unlike the QTS formula
(\ref{eq10}), is manifest local.  In fact the difference between the QTS
formula and the proposed one resides in a quantum fluctuation.  To see this,
consider
\begin{eqnarray} 
<\delta (x)|p|>^2
&=&
\sum_{n,n'}
<n|e^{-\beta H} \delta (x) |p| 
\nonumber\\
&\times&
\left[
\frac{|n><n'|e^{-\beta H} \delta (x)}{Z_0}
\right]
|p||n'>/Z_0
\nonumber
\end{eqnarray}
where the plane wave complete set may be used for $\{n\}$.  Neglecting a
quantum fluctuation, the term $[...] \to \delta_{n~n'} <n|\delta (x) |n'>$, we
reach
\begin{equation} \label{eq18}
<\delta (x) |p|>^2 
~~\longrightarrow~~
<\delta (x) p^2>/L
\end{equation}

We have tested the proposed formula (\ref{eq16}) versus the QTS formula
(\ref{eq3}) for the $I-V$ characteristic of the single Josephson junction
(JJ).
The resistively shunted and current biased single JJ is described by the
action for the superconducting phase difference between the two superconductors\cite{Ambegaokar},
\begin{eqnarray}
S
&=& 
\int_0^{\beta\hbar} d\tau
\left[\frac{C}{2}\left(\frac{\hbar}{2e}\dot{\phi}
\right)^2 + U(\phi)\right]
\nonumber\\
& &
-\int_0^{\beta\hbar}
\int_0^{\beta\hbar}
d\tau d\tau'
\alpha(\tau-\tau') 
{\rm cos}
\left(
\frac{\phi(\tau)-\phi(\tau')}{2}
\right)
\nonumber
\end{eqnarray}
\begin{eqnarray} \label{eq19}
U(\phi)=
-E_{J}{\rm cos}\phi-\frac{I\hbar}{2e}\phi
\nonumber\\
\alpha(\tau)=
\frac{\hbar}{2\pi e^2R_{\rm T}}
\frac{(\pi/\beta\hbar)^2}
{{\rm sin}^2(\pi\tau/\beta\hbar)}
\end{eqnarray}
where $C$ is the capacitance, $R_T$ the shunt resistance, $I$ the bias
current and $E_J$ the Josephson energy.  Thermal over-the-hill motion or
quantum tunneling of the phase gives rise to a voltage $V=\hbar\dot{\phi}/2e$
(the Josephson relation) with $\dot{\phi}=[\phi,H]/i\hbar=(2e)^2n/\hbar C$
where $n=-i\partial_\phi$ is the number of the Cooper pairs.  So the
flux operator here is 
\begin{equation}
{\cal F} 
=
\delta (\phi-\phi_0) 
\frac{(2e)^2}{\hbar C}n
\nonumber
\end{equation}
where $\phi_0$ denotes a maximum point of the potential $U(\phi)$. 
Taking into account the backward flux by a detailed balance, we have
\begin{eqnarray} \label{eq20}
\frac{V}{e/2C}
=
\left[1-{\rm exp}(-\pi I\hbar\beta/e)\right]\times\frac{1}{2}
\sqrt{
\langle\delta(\phi-\phi_0)n^2/L\rangle
}
\end{eqnarray}
After some manipulations, one can express the average in (\ref{eq20}) as
\begin{eqnarray} \label{eq21}
\langle\delta(\phi-\phi_0)n^2\rangle
=
Z/Z_0
\nonumber\\
Z
=
-\frac{1}{4}\partial_{\phi}^{2}W(\phi,2\phi_0-\phi)
\arrowvert_{\phi~=~\phi_0}
\nonumber\\
Z_{0}
=
\int d\phi W(\phi,\phi)
\nonumber\\
W(\phi,\phi')
=
\int_{\phi\to\phi'}{\cal D}
\tilde{\phi}~{\rm exp}(-S/\hbar)
\end{eqnarray}
where $\tilde{\phi}(0)=\phi$ and $\tilde{\phi}(\beta\hbar)=\phi'$.

The path-integral (\ref{eq21}) with the action $S$ given by (\ref{eq19}) can be
evaluated precisely by the cluster transfer matrix (CTM) method\cite{Chung1}.  In the present problem, however, the dimensionless
junction conductance $g\equiv R_q/R_T$ where
$R_Q=h/4e^2=6.45~K\Omega$ is the quantum resistance, is at $g\lesssim1$,
and from the study of the single electron box which has a similar action as
(\ref{eq19}), this regime of $g$ can be accurately handled by the 1-cluster TM
method\cite{Chung2}. 

For a fixed g, the temperature
dependence of the resistance changes from insulator-like behavior
$\frac{dR}{dT}~<~0$ to superconductor-like $\frac{dR}{dT}~>~0$ with incresing
ratio $E_J/E_C$ where $E_C\equiv \frac{e^2}{2C}$.  Repeating the calculations for different $g$, we can thus
map a superconductor-insulator (SI) phase diagram in the $E_J/E_C - g$ plane.
Fig.~1 shows a SI phase diagram at T=80 mK.  The corresponding experimental
results are denoted by open circles (S-like) and solid circles
(I-like)\cite{Hakonen,Yagi}.  Our result is the open diamond, the band-theory
result the thick solid line\cite{Zorin}, and the QTS result is denoted by triangles.
While there is an issue of temperature dependence concerning the phase
diagram\cite{Chung2}, the agreement between the previous theory and
experiments with the proposed method, not with the QTS theory, will be
evident. A major disagreement between theory and experiment
is for some data points near $g=2.8$\cite{Yagi}.  However a similar phase
diagram experimentally found for the 2D JJ arrays with similar
parameter ranges for $g, E_{j}/E_{C}$ and $T$ is bounded, $E_{J}/E_{C}\lesssim
0.5$, and $g\lesssim 0.5$ (cf. Fig~3 in\cite{Yamaguchi}).  The above data near $g=2.8$
is currently mysterious.

In conclusion, we have proposed a new formula for calculating the quantum
Kramers rate which may correspond to the Landauer formula for electrical
conductance.  The single Josephson junction for which the proposed formula was
tested is quite general: It contains quantum fluctuation, dissipation and
external bias.  The reasonable outcome of the proposed formula in comparison
with experiments and the previous theory, although the latter is not free from
criticism\cite{Chung2}, may be encouraging a further testing of the proposed
method in a variety of physical systems. On the other hand, we must also admit
that the new formula was derived in an attempt for technical improvement
over the QTS theory.  A contemplation of deep physical argument leading directly
to the proposed formula is much desired.

This work was supported by the Visitor Program of the MPI-PKS. The work was
also partially supported
 by the NSF under Grant 
No. DMR990002N and utilized the SGI/CRAY Origin2000 at the National Center 
for Supercomputing Applications at the University of Illinois at
Urbana-Champaign. I thank Yoshiro Kakehashi, Mitsuhiro Arikawa and Klaus
Morawetz for helpful discussions and bringing important references to my
attention.
I also thank Peter H\"anggi and Anthony Leggett for enlightening conversations.

\begin{figure} 
\epsfig{file=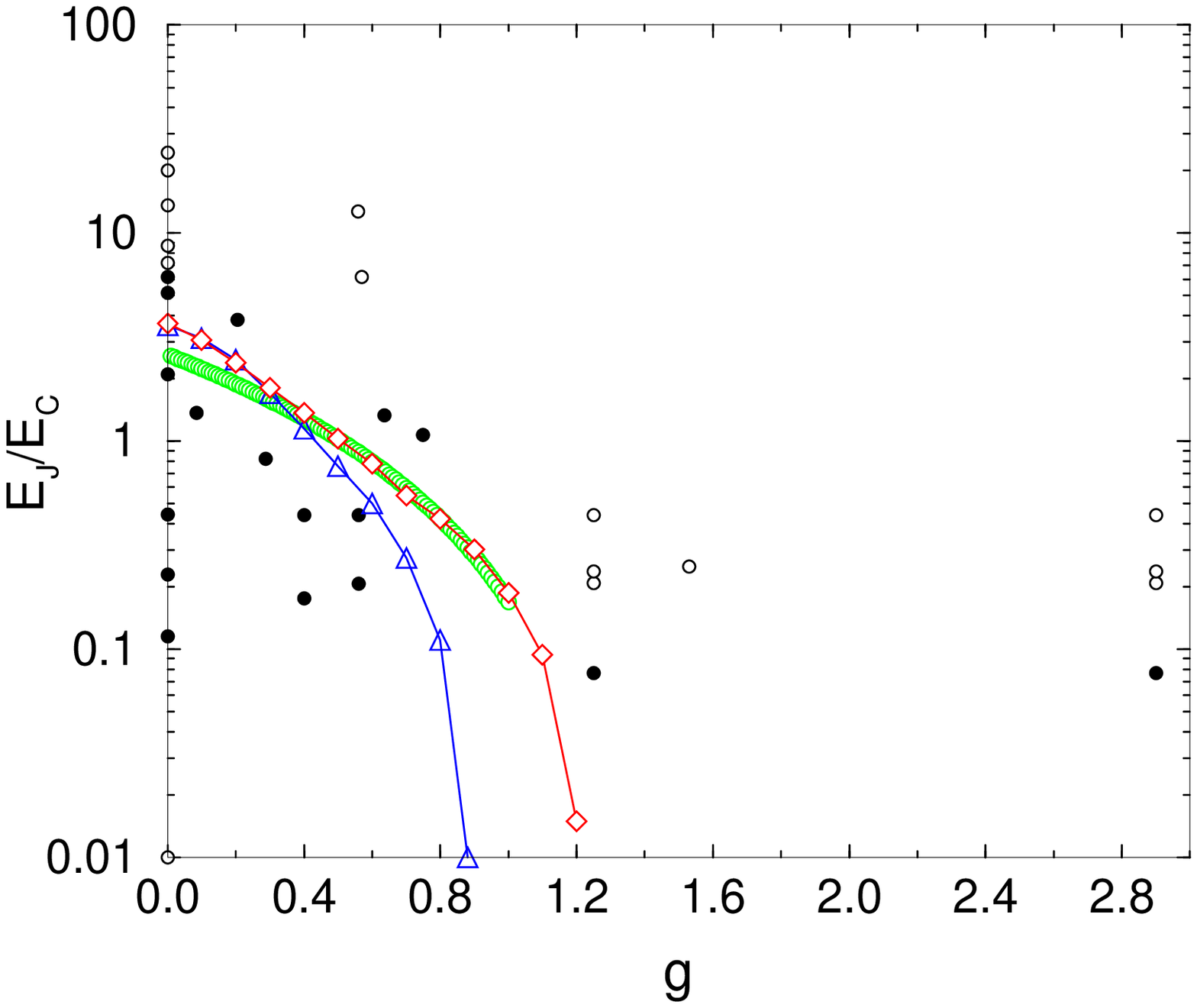,width=6.5cm,height=7.2cm,angle=-90}
\end{figure}
\vspace{-0.5cm}
FIG.~1. Phase diagram of the single Josephson junction. 
The phase boundary lies
between the insulator-like (solid circles) and 
superconductor-like (open
circles) samples experimentally found in [14,15].  
The thick line
is the band theory [16] at $T=80~mK$. The triangle is 
the QTS theory and 
diamond is due to the present theory at $T=80~mK$.

\end{multicols}

\end{document}